\documentclass{aa} %
\usepackage{graphics}
\usepackage{natbib}
\bibpunct{(}{)}{;}{a}{}{,}
\usepackage{txfonts}
\begin{document}
\title{Resonance-Enhanced Two-Photon Ionization (RETPI) of \ion{Si}{ii} and an Anomalous, Variable Intensity 
of the $\lambda$1892 \ion{Si}{iii}] Line in the Weigelt Blobs of Eta Carinae}
\titlerunning{RETPI of \ion{Si}{ii} and anomalous $\lambda$1892}
\authorrunning{Johansson et al.}
\author{S. Johansson\inst{1} \and H. Hartman \inst{1} \and  V.S. Letokhov\inst{2,1}}
\institute{Lund Observatory, Lund University, P.O. Box 43, SE-22100, Lund, SWEDEN \and Institute of Spectroscopy, 
Russian Academy of Sciences, 142190 Troitsk, Moscow region, RUSSIA}
\date{Received <date> / Accepted <date>}

\abstract{The Si III] 1892 \AA\ intercombination line shows an anomalously high 
intensity in spectra of the radiation-rich Weigelt blobs in the vicinity of Eta Carinae. The 
line disappears during the 100 days long spectral events occurring every 5.5 
years.
The aim is to investigate whether resonance-enhanced 
two-photon ionization (RETPI) is
a plausible excitation mechanism for the Si III] $\lambda$1892 line.
The possible intensity enhancement of the $\lambda$1892 line is investigated
as regards quasi-resonant intermediate energy levels of Si II.
The RETPI mechanism is effective on \ion{Si}{ii} in the radiation-rich Weigelt 
blobs where the two excitation steps are provided by the two intense hydrogen lines 
Ly$\alpha$ and Ly$\gamma$.
\keywords{atomic processes --
radiation mechanisms: non-thermal: two-photon ionization -- HII regions -- stars: 
individual: Eta Carinae -- Line: formation}}

\maketitle 
\section{Introduction}

\citet{JL01b} have considered the possibility that ions 
can be created by resonance-enhanced two-photon ionization (RETPI) in a low-density 
($N_{\mathrm H} < 10^{10}\, \mathrm{cm}^{- 3}$) astrophysical plasma, in which 
the HLy$\alpha$ line  is supposed to have a 
sufficiently high intensity. In subsequent papers, this possibility was examined 
for the elements C, N, O 
\citep{JL01d,JL02} and the rare gases Ne, 
Ar 
\citep{JL04b}, with the involvement of intense 
H Ly$\alpha, \beta, \gamma $ as well as HeI and HeII lines.

The RETPI process can compete with ionization by collisions between atoms (ions) 
and free electrons in a radiation-rich astrophysical plasma, in which the 
radiation energy density is comparable with or even 
higher than the energy density of free electrons. Such an astrophysical plasma 
can, for example, be represented by a gas cloud ejected from a hot star. The 
best-known case is the Weigelt blobs (WB's) near one of the most massive and 
brightest stars in the Galaxy, Eta Carinae (HD93308) 
\citep{WE86}. The emission 
line spectra of these blobs have been resolved from the radiation of the central 
star by means of the STIS two-dimensional spectrograph aboard the Hubble Space 
Telescope 
\citep{GID01}. The STIS observations show that the blobs are 
located only a few hundred stellar radii from the central star, and that their 
size and hydrogen density imply a high optical depth in the Lyman continuum 
\citep{DH97,BKS03}.

Simple estimates show that the effective spectral temperature of Ly$\alpha$ 
inside the blobs $T_{\alpha}^{\mathrm{eff}} \simeq (10 \; \mathrm{to} \; 15) 
\cdot 10^3$ K 
\citep{JL04a}. This value is comparable to or 
even higher than the electron temperature, which implies a sharp change of the 
energy balance in the WB's in comparison with classical planetary nebulae. This 
change is explained by the fact that the dilution factor of the radiation 
reaching the blobs from the central star is compensated by the effect of 
"spectral compression". Thus, the Lyman-continuum radiation, which is absorbed 
in the photoionization of hydrogen, is emitted in the relatively narrow  lines 
HLy$\alpha, \beta, \gamma$ as a result of radiative recombination. The WB's  may 
therefore be regarded as \textit{radiation-rich} nebulosities to distinguish 
them from \textit{thermal} planetary nebulae \citep{a84_v2}. 

\begin{figure}
\resizebox{\columnwidth}{!}{\includegraphics{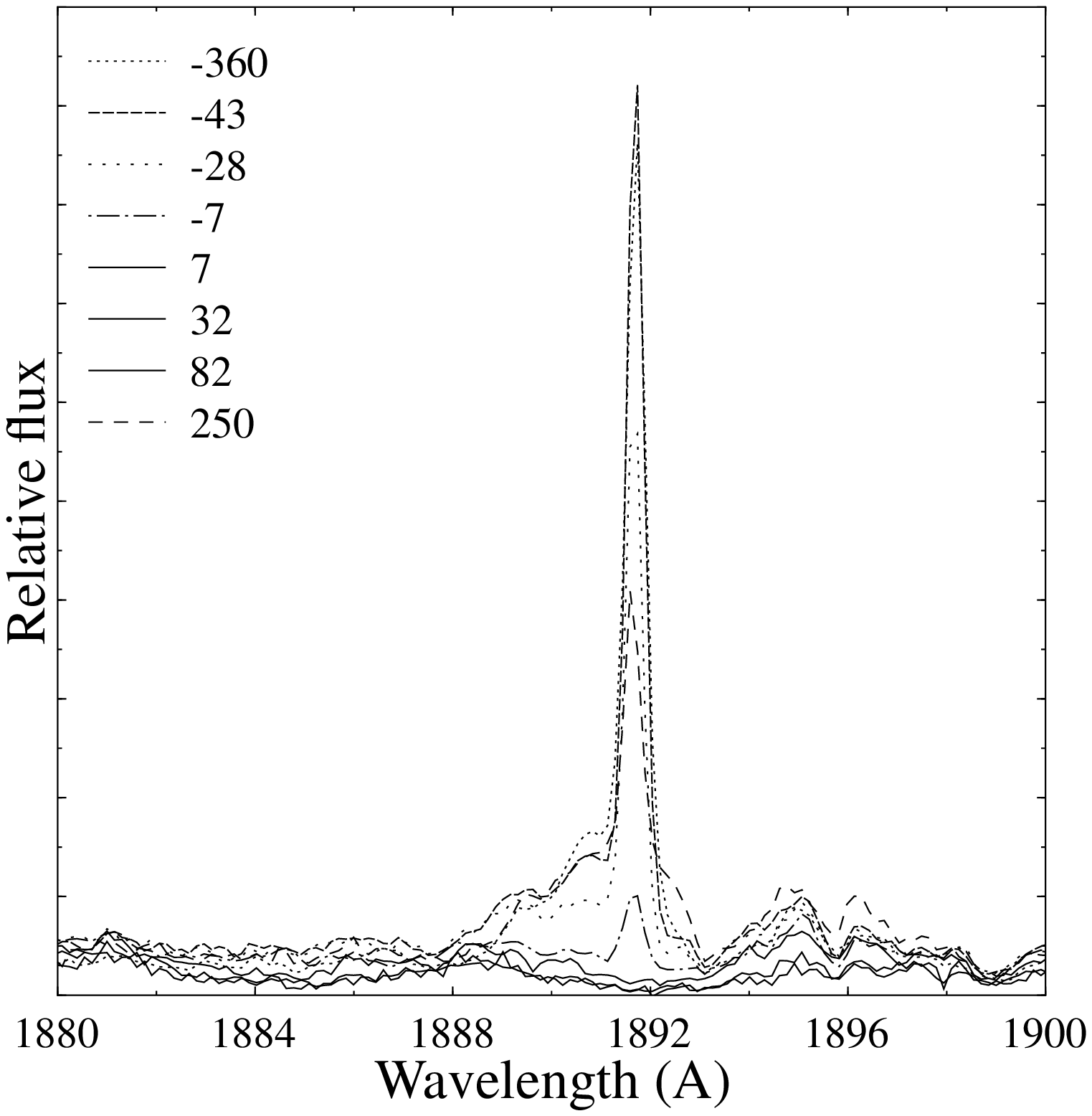}}\\ 
\resizebox{\columnwidth}{!}{\includegraphics{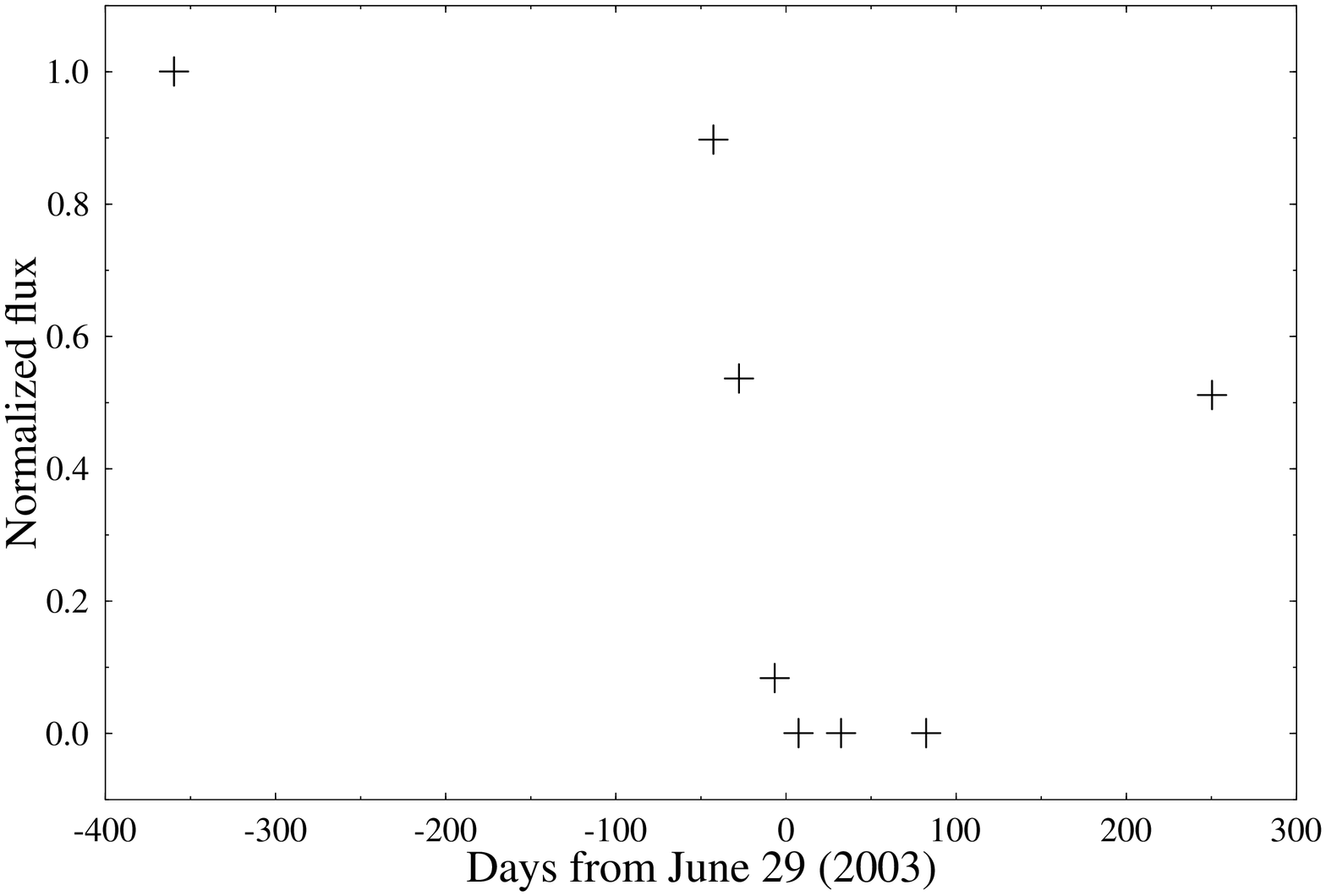}}
\caption{Time variation of the intensity of the anomalous intercombination \ion{Si}{iii} line at 1892 \AA\  
observed in the Weigelt blobs of Eta Carinae during the spectral event 2003. (a) Spectral profiles of the line; 
(b) The intensity change during a period of two years around the event. The data are part of HST Treasury Project 
of Eta Carinae.}
\end{figure}

The WB's represent naturally a suitable astrophysical plasma in which the RETPI 
effect may occur. To verify this fact, one can, in principle, investigate the 
ionization equilibrium among various elements in the blobs aiming at finding 
some indications of the RETPI process. Such indications could be abundance 
depletion or enhancement of some particular ion compared to predictions from an 
ionization equilibrium governed by electronic ionization. Another possibility is 
to investigate spectral anomalies that are apparent for some ions. This approach 
is especially valuable, since RETPI is a purely radiative process depending on 
the intensity of the H~Ly$\alpha, \beta, \gamma$  radiation. At the same time, 
there is the well-known "spectral event" in $\eta$ Car, a periodic attenuation 
of the intensity of some lines occurring every 5.5 years and having a duration 
of about three months 
\citep{D96}. For many of these lines it has been 
shown that their intensities depend on H~Ly$\alpha, \beta, \gamma$ radiation. 
Thus, data recorded during these three months allow for a discrimination of the 
excitation mechanism of the spectral lines in the WB's 
\citep{HDJ05}. 
In general, the radiatively excited lines should vanish during the spectral 
event, whereas the recombination lines should remain, as the recombination time 
in the HI zone of the WB's exceeds three months. 

In the present paper, following the above-described strategy of finding proofs of the 
existence of the RETPI effect, we propose an explanation of the spectral anomaly 
of the \ion{Si}{iii}] intercombination line in spectra of the Weigelt blobs.

\section{Striking Spectral Feature: The 1892 \AA\ \ion{Si}{iii}] Intercombination Line}

The emission lines of various elements in the WB's were summarized by 
\citep{zphd01}. In particular, he noted that in 1999 the \ion{Si}{iii}] 3s$^2$ $^1$S$_0$--3s3p 
$^3$P$_1$ intercombination line at 1892 \AA\ was the third strongest emission 
feature in the satellite UV region of the observed spectrum (only surpassed by 
the 2507 and 2509 \AA\ FeII fluorescence lines). However, there is no sign of 
the \ion{Si}{iii}] line in the data recorded during the spectral event in 1998. This is 
perhaps the most striking example of the influence that the spectral event 
imposes on the WB spectrum 

The same effect was observed during the spectral event of 2003. Figures 1a and b 
show the drop of the intensity of the  \ion{Si}{iii}] $\lambda$1892 line according to 
the $HST$/STIS CCD-data of the Weigelt blobs taken during the event in June 
2003, as part of the HST Treasury Project of Eta Carinae. The slit of 
the spectrograph was centered on Weigelt blob D, which is the main contributor 
to the observed line emission. For some position angles of the slit, the 
adjacent blobs B and C are not fully excluded from the field of view. Thus, B 
and C also contribute to the observed flux, but this contribution does not 
change the general behavior of the curves in Figs. 1a and 1b. The rate of the 
intensity decrease for the \ion{Si}{iii}] line coincides approximately with that of 
twelve FeII spectral lines excited by Ly$\alpha$ radiation 
\citep{HDJ05}. This fact suggests that the excitation of the \ion{Si}{iii}] $\lambda$1892 line 
should also be associated with Ly$\alpha$ radiation.

\begin{figure}
\resizebox{\columnwidth}{!}{\includegraphics{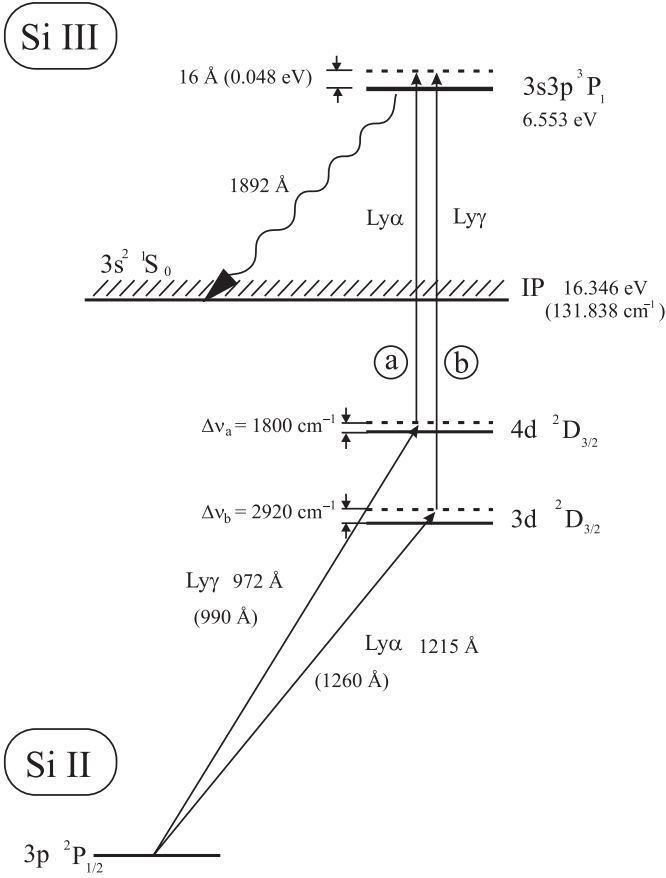}}
\caption{Two possible pathways of resonance-enhanced two-photon ionization (RETPI) of \ion{Si}{ii} by HLy$\alpha$ 
and HLy$\gamma$ radiation providing excitation of the 1892 \AA\ \ion{Si}{iii}] intercombination line.}
\end{figure}

\section{Scheme of RETPI operating on \ion{Si}{ii} and Involving Excitation of the \ion{Si}{iii}] 
Intercombination Line}

The ionization potential of \ion{Si}{i} is 8.12 eV, i.e. it is lower than the ionization 
potential of HI. Therefore, ionization of \ion{Si}{i} can be achieved by means of 
stellar Planck radiation in the spectral range 912\AA\ $< \lambda <$1527 \AA\ 
that penetrates inside the WB. The RETPI of \ion{Si}{ii} to form \ion{Si}{iii} can occur as a 
result of two-photon absorption of H Ly$\alpha$ radiation 
\citep{JL01d} But in that case, the excess of energy provided by the two 
Ly$\alpha$ photons after ionizing Si II to Si III, $2 \mathrm h 
\nu$(Ly$\alpha$)- IP(Si II), amounts to 4.06 eV. This is insufficient to 
populate the 3s3p~$^3$P$_1$ state of \ion{Si}{iii} whose excitation energy is 6.553 eV. 
However, the RETPI of \ion{Si}{ii} under the effect of a combination of two different 
Lyman lines, Ly$\alpha$ and Ly$\gamma$, provides for the excitation of a state 
in the Si II continuum at an energy of E = 6.608 eV. This is E = 0.055 eV higher than 
the excitation energy of the triplet state. Based on this fact two possible 
pathways of the RETPI of \ion{Si}{ii} are illustrated in Figure 2.

The coupling between the excited state in the \ion{Si}{ii} continuum and the 
3s3p~$^3$P$_1$ triplet state of \ion{Si}{iii} may prove quite enough for its excitation, 
followed by a radiative transition to the ground state 
in \ion{Si}{iii}, i.e. emission of the  \ion{Si}{iii}] $\lambda1892$ intercombination line. This 
is the only allowed radiative decay channel of the 3s3p~$^3$P$_1$ state, and 
because of a relatively strong LS coupling the transition probability of this 
LS-forbidden lines is A$=1.67 \cdot 10^4$ s$^{-1}$ \citep{KJS83}.

Considering that the Ly$\alpha$ and Ly$\gamma$ spectral lines are generated in 
the HII zone of the stellar wind as well, the energy difference in the 
excitation of the triplet state is even smaller than  E = 0.055 eV ($\approx$16 \AA). 
With the terminal velocity  $v_{\mathrm{term}}$ of the stellar wind from Eta 
Carinae being as high as +625 km~s$^{-1}$ 
\citep{HDI01,SDG03} 
the two photons, Ly$\alpha$ and Ly$\gamma$, irradiating the WB from the opposite 
side, reduce the difference to 10 \AA, which increases the coupling between the 
continuum states of \ion{Si}{ii} and the triplet state of \ion{Si}{iii}.

Energetically, the RETPI process proposed could also populate the J=0 and J=2 levels
of the 3s3p~$^3$P term, yielding an energy difference E=0.022 eV ($\approx$6 \AA) for the
J=2 level. However, the radiative decay of this level involves a forbidden transition at 
1882.7 \AA\ whose gA-value is 6 orders of magnitude smaller than the value
for to the observed $\lambda1892$ 
intercombination line. The forbidden $\lambda1882$ line is not observed, which could partly
be due to a collisional ion-electron coupling between the J=2 and J=1 levels.

\section{Rate of the RETPI of \ion{Si}{ii} by the Ly$\alpha$ and Ly$\gamma$ Radiation}
The rate $W_{1i} (\mathrm{s}^{-1})$ of the RETPI of \ion{Si}{ii} under the effect of the 
Ly$\alpha$ + Ly$\gamma$  two-frequency radiation for each of the pathways, (a) 
and (b), of Fig. 2 is defined by the following expressions \citep{JL01d}:

\begin{equation}
 W_{1i}^a \simeq \frac{2}{\pi} \cdot 
\frac{\delta \nu_\alpha \delta \nu_\gamma}{(\Delta \nu_a)^2} \cdot 
\frac{\sigma_{2i}^a}{\lambda_\alpha^2} 
A_{21}^a \mathrm {exp} \left ( -\frac{h \nu_\gamma}{kT_\gamma^{\mathrm {eff}}}\right )
 \mathrm {exp} \left (- \frac{h \nu_\alpha}{kT_\alpha^{\mathrm {eff}}} \right )                          
\end{equation}

and

\begin{equation} 
W_{1i}^b \simeq \frac{2}{\pi} \cdot 
\frac{\delta \nu_\alpha \delta \nu_\gamma}{(\Delta \nu_b)^2} \cdot 
\frac{\sigma_{2i}^b}{\lambda_\gamma^2} 
A_{21}^b \mathrm {exp} \left ( -\frac{h \nu_\alpha}{kT_\alpha^{\mathrm {eff}}} \right )
\mathrm {exp} \left ( - \frac{h \nu_\gamma}{kT_\gamma^{\mathrm {eff}}} \right ) 
\end{equation}

where $\delta \nu_\alpha$ and $\delta \nu_\alpha$ are the spectral widths of the 
Ly$\alpha$ and Ly$\gamma$ lines, $\Delta \nu_a$ = 1800 cm$^{-1}$ and $\Delta 
\nu_b$ = 2920 cm$^{-1}$ are the frequency detunings of the Ly$\alpha$ and 
Ly$\gamma$ lines relative to the 4d~$^2$D$_{3/2}$ and 3d~$^2$D$_{3/2}$  
intermediate quasi-resonant levels, respectively. $\lambda_\alpha$ and 
$\lambda_\gamma$ are the wavelength of the Ly$\alpha$ and Ly$\gamma$ lines, 
$\sigma_{2i}^a$ and   $\sigma_{2i}^b$ the cross-sections for photoionization 
from the 4d~$^2$D and 3d~$^2$D,  $A_{21}^a$ and $A_{21}^b$ the 
Einstein coefficients for the 4d~$^2$D $\rightarrow$ 3p~$^2$P and 3d~$^2$D 
$\rightarrow$ 3p~$^2$P  radiative transitions,  
and $T_\alpha^{\mathrm {eff}}$ and $T_\gamma^{\mathrm {eff}}$ are the effective 
(spectral) temperatures of the Ly$\alpha$ and Ly$\gamma$ radiation, 
respectively. In the scheme in Fig.2 we have only included parameter values 
for the ground state transitions, 3p~$^2$P$_{1/2}$ 
$\rightarrow$ 4d~$^2$D$_{3/2}$  and  3p~$^2$P$_{1/2}$ $\rightarrow$ 
3d~$^2$D$_{3/2}$, but there are also contributions from the 3/2$\rightarrow$5/2 
and 3/2$\rightarrow$3/2 fine structure transitions. 

Let us make a very approximate estimate of the rates $W_{1i}^a$ and $W_{1i}^b$ 
of the RETPI of \ion{Si}{ii} under the effect of the Ly$\alpha$ and Ly$\gamma$ radiation 
with the spectral widths $\delta \nu_\alpha \simeq \delta \nu_\gamma \simeq 
500$~cm$^{-1}$, which corresponds to the widths of these lines in the HII 
region of the stellar wind in the vicinity of the WB. Assuming approximately 
that $A_{21} \simeq 10^9$~s$^{-1}$ and $\sigma_{2i} \simeq 10^{-
18}$~cm$^{2}$, we obtain the following estimate:

\begin{equation}
 W_{1i}^a \simeq 
W_{1i}^b \simeq  0.2 \cdot \mathrm {exp} \left (
-\frac{h \nu_\alpha}{kT_{\mathrm {eff}}^\alpha} \right )
\mathrm {exp} \left ( 
-\frac{h \nu_\gamma}{kT_{\mathrm {eff}}^\gamma} \right ) \; \; \; [ \mathrm{s}^{-1} ] 
\end{equation}

\begin{figure}
\resizebox{\columnwidth}{!}{\includegraphics{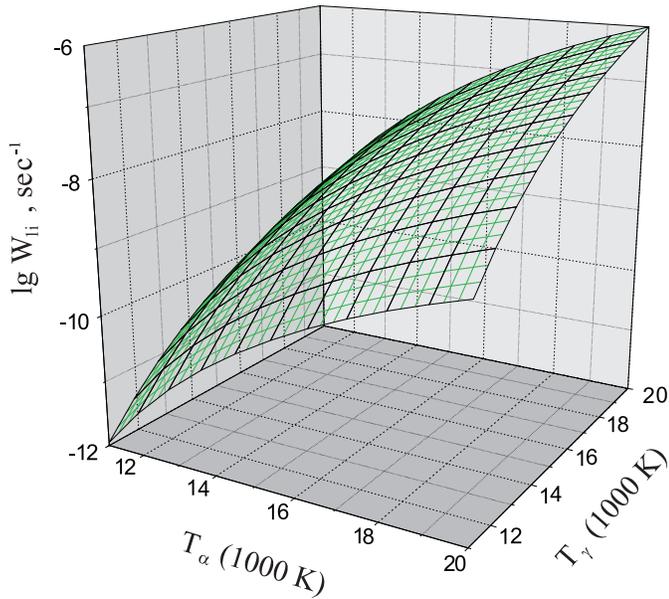}}
\caption{Rate of RETPI (s$^{-1}$) of \ion{Si}{ii} as a function of the effective (spectral) temperatures 
of Ly$\alpha$ and Ly$\gamma$.}
\end{figure}

Figure 3 presents the relationship between the total rate of the RETPI process 
involving the two pathways, $W_{1i}=W_{1i}^a+W_{1i}^b$, and the 
effective temperatures $T_\alpha^{\mathrm {eff}}$ and $T_\gamma^{\mathrm {eff}}$ 
in the range (10 -- 20)$\cdot10^3$ K. The approximate estimate of 
the rate of the RETPI of \ion{Si}{ii} accompanied by the excitation into its ionized 
continuum close to the triplet state of \ion{Si}{iii} lies in the range $10^{-
9}-10^{-6}$ s$^{-1}$.
	
To use this estimate to get an explanation of the intensity observed for the \ion{Si}{iii}] 
$\lambda$1892 line (Fig. 1) seems rather natural. However, such an 
estimate would be rather approximate, since the volumes of the WB's and stellar 
wind regions wherein this intercombination line is generated by the 
RETPI process are unknown. The abundance of Si in 
the WB is only approximately known, as is the degree of coupling between the 
triplet state of \ion{Si}{iii} and the ionization continuum of \ion{Si}{ii}. Nevertheless, one 
can note that with the total volume of the WB and the surrounding stellar wind 
region being $V \simeq 10^{47}$~cm$^3$, the Si abundance 
$N_{\mathrm {Si}}=5 \cdot 10^{-5} N_\mathrm H$, with $N_\mathrm H \simeq 
10^8$~cm$^{-3}$, and the coupling constant of the order of unity, the \ion{Si}{ii} RETPI 
rate  $W_{\mathrm {1i}} \simeq 10^{-7}-10^{-8}$~s$^{-1}$ explains fairly well 
the observed intensity of the 1892 \AA\ \ion{Si}{iii} line, which from 
the blobs is 7$\cdot$10$^{-12}$ erg\,cm$^{-2}$\,s$^{-1}$\,\AA $^{-1}$ \citep{GID01}.

\section{Conclusion}

This is the first attempt to use the rich observational data on emission line 
spectra of the radiation-rich Weigelt blobs in the vicinity of Eta Carinae and 
search for evidence for resonance-enhanced two-photon ionization (RETPI) in an 
astrophysical object. This object is special for this purpose for several 
reasons. Firstly, the availability of spectral data from $HST$/STIS with 
excellent spatial resolution (no overlap with the radiation from the central 
star). Secondly, the effect of a periodical reduction of the ionizing radiation 
from the central star during 'spectroscopic events' allows to distinguish the 
radiative (collisionfree) and recombination (collisional) excitation mechanisms 
\citep{HDJ05}. It is rather tempting to search for other  anomalies in 
the spectra of the Weigelt blobs to reveal possible contributions from the RETPI 
process in the ionization of elements.

\begin{acknowledgements} 
We thank the referee Prof.~C.~Jordan for very useful comments on the manuscript.
This research project is supported by a grant (S.J.) 
from the Swedish National Space Board. V.S.L. acknowledges the financial support 
through grants (S.J.) from the Wenner-Gren Foundations, as well as Lund Observatory 
for hospitality, and the Russian Foundation for Basic Research (Grant No 03-02-16377). 
This work is based on observations made with the NASA/ESA Hubble Space 
Telescope, proposal IDs: 9337, 9420 and 9973. STScI is operated by the 
Association of Universities for Research in Astronomy, Inc.\ under NASA contract 
NAS 5-26555. We are grateful to Dr.\ Theodore Gull for providing reduced and 
calibrated spectra.
\end{acknowledgements} 
  
\bibliographystyle{aa}
\bibliography{johansson}

\begin{thebibliography}{16}
\expandafter\ifx\csname natexlab\endcsname\relax\def\natexlab#1{#1}\fi

\bibitem[{{Aller}(1984)}]{a84_v2}
{Aller}, L.~H. 1984, {Physics of thermal gaseous nebulae} ({D.\ Riedel Publ.\
  Co., Dordrecht})

\bibitem[{{Damineli}(1996)}]{D96}
{Damineli}, A. 1996, ApJL, 460, L49

\bibitem[{{Davidson} \& {Humphreys}(1997)}]{DH97}
{Davidson}, K. \& {Humphreys}, R.~M. 1997, ARA\&A, 35, 1

\bibitem[{{Gull} {et~al.}(2001){Gull}, {Ishibashi}, {Davidson}, \&
  {Collins}}]{GID01}
{Gull}, T., {Ishibashi}, K., {Davidson}, K., \& {Collins}, N. 2001, in ASP
  Conf. Ser. 242: Eta Carinae and Other Mysterious Stars: The Hidden
  Opportunities of Emission Spectroscopy, 391

\bibitem[{{Hartman} {et~al.}(2005){Hartman}, {Damineli}, {Johansson}, \&
  {Letokhov}}]{HDJ05}
{Hartman}, H., {Damineli}, A., {Johansson}, S., \& {Letokhov}, V.~S. 2005,
  \aap, 436, 945

\bibitem[{{Hillier} {et~al.}(2001){Hillier}, {Davidson}, {Ishibashi}, \&
  {Gull}}]{HDI01}
{Hillier}, D.~J., {Davidson}, K., {Ishibashi}, K., \& {Gull}, T. 2001, ApJ,
  553, 837

\bibitem[{{Johansson} \& {Letokhov}(2001{\natexlab{a}})}]{JL01b}
{Johansson}, S. \& {Letokhov}, V. 2001{\natexlab{a}}, Science, 291, 625

\bibitem[{{Johansson} \& {Letokhov}(2001{\natexlab{b}})}]{JL01d}
{Johansson}, S. \& {Letokhov}, V.~S. 2001{\natexlab{b}}, \aap, 375, 319

\bibitem[{{Johansson} \& {Letokhov}(2002)}]{JL02}
{Johansson}, S. \& {Letokhov}, V.~S. 2002, \aap, 395, 345

\bibitem[{{Johansson} \& {Letokhov}(2004{\natexlab{a}})}]{JL04a}
{Johansson}, S. \& {Letokhov}, V.~S. 2004{\natexlab{a}}, Astronomy Letters, 30,
  58

\bibitem[{{Johansson} \& {Letokhov}(2004{\natexlab{b}})}]{JL04b}
{Johansson}, S. \& {Letokhov}, V.~S. 2004{\natexlab{b}}, Astronomy Reports, 48,
  399

\bibitem[{{Kwong} {et~al.}(1983){Kwong}, {Johnson}, {Smith}, \&
  {Parkinson}}]{KJS83}
{Kwong}, H.~S., {Johnson}, B.~C., {Smith}, P.~L., \& {Parkinson}, W.~H. 1983,
  \pra, 27, 3040

\bibitem[{{Smith} {et~al.}(2003){Smith}, {Davidson}, {Gull}, {Ishibashi}, \&
  {Hillier}}]{SDG03}
{Smith}, N., {Davidson}, K., {Gull}, T.~R., {Ishibashi}, K., \& {Hillier},
  D.~J. 2003, ApJ, 586, 432

\bibitem[{{van Boekel} {et~al.}(2003){van Boekel}, {Kervella}, {Sch{\" o}ller},
  {Herbst}, {Brandner}, {de Koter}, {Waters}, {Hillier}, {Paresce}, {Lenzen},
  \& {Lagrange}}]{BKS03}
{van Boekel}, R., {Kervella}, P., {Sch{\" o}ller}, M., {et~al.} 2003, \aap,
  410, L37

\bibitem[{{Weigelt} \& {Ebersberger}(1986)}]{WE86}
{Weigelt}, G. \& {Ebersberger}, J. 1986, A\&A, 163, L5

\bibitem[{Zethson(2001)}]{zphd01}
Zethson, T. 2001, PhD thesis, Lund University

\end{thebibliography}

\end{document}